\documentclass[a4paper,fleqn,usenatbib]{mnras} %\pdfoutput=1
\usepackage{graphicx}
\usepackage{graphics}
\usepackage{amsmath, amssymb}
\usepackage{multirow}
\usepackage{color}
\usepackage{xcolor}
\usepackage{bm}		% Bold maths symbols, including upright Greek
\usepackage[normalem]{ulem}

\usepackage{float}

\newcommand{\feh} {\mbox{\rm [Fe/H]}}

\newcommand{\Vphi} {\mbox{\rm $V_{\rm{\phi}}$}}
\newcommand{\kmprs} {\mbox{\rm \,$km/s$\,}}
\newcommand{\kG} {\mbox{\rm \,$kpc/Gyr$\,}}

\title{Open Clusters as Tracers on Radial Migration of the Galactic Disk}

\author[Chen \& Zhao]{Y.Q. Chen \& G. Zhao \thanks{E-mail:cyq,gzhao@nao.cas.cn}
\\
$^{1}$ Key Laboratory of Optical Astronomy, National Astronomical Observatories, Chinese Academy of Sciences, Beijing, 100101, China \\
$^{2}$ School of Astronomy and Space Science, University of Chinese Academy of Sciences, Beijing 101408, China
}

\date{Submitted Feb. 2020}

\pubyear{0000}

\begin{document}
\label{firstpage}
\pagerange{\pageref{firstpage}--\pageref{lastpage}}
\maketitle

\begin{abstract}
Radial migration is an important process in the Galactic disk.
A few open clusters show some evidence on this mechanism but
there is no systematic study.
In this work, we investigate the role of radial migration 
on the Galactic disk based on a large sample of 146 open clusters
with homogeneous metallicity and age from Netopil et al. and
kinematics calculated from \textit{Gaia} DR2.
The birth site $R_b$, guiding radius $R_g$ and other
orbital parameters are calculated, and the
migration distance $|R_g-R_b|$ is obtained, 
which is a combination of metallicity, kinematics 
and age information.
It is found that 44\% open clusters have $|R_g-R_b|< 1$ kpc,
for which radial migration (churning) is not significant.
Among the remaining 56\% open clusters with $|R_g-R_b|> 1$ kpc, 
young ones with $t<1.0$ Gyr tend to migrate inward, while older clusters
usually migrate outward.
Different mechanisms of radial migration between young and old
clusters are suggested based on their different migration rates, 
Galactic locations and orbital parameters. 
For the old group, we propose a plausible way to estimate migration rate
and obtain a reasonable value of $1.5 \pm 0.5 \kG$ based on ten
intermediate-age clusters at the outer disk,
where the existence of several special clusters implies its complicate
formation history.
\end{abstract}

\begin{keywords}
%\textit{Unified Astronomy Thesaurus concepts:} Milky Way dynamics (1051); Milky Way evolution (1052); Milky Way disk (1050); Open Clusters (1000)
Galaxy: disc; Galaxy: evolution; Galaxy: abundances; Galaxy: kinematics and dynamics; open clusters and associations: general
\end{keywords}

%%%%%%%%%%%%%%%%%%%%%%%%%%%%%%%%%%%%%%%%%%%%%%%%%%

%%%%%%%%%%%%%%%%% BODY OF PAPER %%%%%%%%%%%%%%%%%%

\section{Introduction}
\label{sect:introdisc}
Radial migration of stars in the Galactic disk is an important process
as shown by observational works \citep{Grenon72, Wielen96, 
Haywood08, Hayden15} and theoretical models/simulations \citep{SB02,
Roskar08, Minchev13}.
It has been invoked to justify many observational phenomena, such as
the existence of metal-rich stars in the solar
neighbourhood \citep{Casagrande11, Minchev13}, the substantial scatter
in the age-metallicity relation \citep{Edvardsson93,Haywood13}, the
skewness of the metallicity distributions \citep{Hayden15},
the presence
of low-velocity dispersion of [$\alpha$/Fe]-enhanced stars
in the solar neighborhood \citep{Minchev14}.

In particular, the lack of age-metallicity relation
is contrary to observations of interstellar gas and young stars
in galaxies that stars born at the same epoch have very similar
[Fe/H] \citep{Przybilla08,Nieva12}.
% (e.g., Przybilla et al. 2008,Nieva \& Pryzbilla 2012).
A solution to the discrepancy is radial migration, 
first proposed by \cite{SB02}. They suggested that stars and gas 
close to the co-rotation resonance
(associated with the spiral structure) experience large changes in
their positions by migrating outward and inward radially.
Subsequently, more analytic and simulation
works \citep{Roskar08,Minchev13,Loebman16} recognized that
migration of stars (or clusters) from their birth radius of inner disk
to the solar neigborhood
significantly broadens local age and metallicity distributions.
Besides spiral arms, minor satellites \citep{Quillen09}
and the bar structure \citep{Minchev10} are also
radial migration mechanisms. In particular,
the coupling between spiral arms and bar structure \citep{Minchev11,Minchev13}
is found to invoke a very powerful stellar radial migration
in the inner disk, while perturbations caused by minor mergers
are more effective in the outer disks \citep{Quillen09,Bird12}.

Radial migration could alter the disk structure and
chemical composition of local regions via the churning process.
According to \cite{SB02}, churning
occurs in the presence of changing and complex
non-axisymmetric patterns (over-densities) such as
spiral arms, which exert torques on stars and lead to an
effective change in a star's angular momentum (or guiding
radius). By conserving their angular momentum,
blurring refers to orbital heating
by a variety of perturbations in the in-plane or vertical
direction, which may cause increasing epicycles but
remain radial oscillations around the guiding radii.
Usually, churning has a stronger effect than blurring on
the Galactic disk because it can migrate stars by several kpc
over time-scales as short as a few Gyr \citep{Kubryk13,Grand16}.
In this work, radial migration refers to the churning process,
although some works \citep{Halle15,Halle18} include
both churning and blurring in the nomenclature of
radial migration.

It is important to find birth sites of stars and derive
migration distances in order to investigate
its effect on the Galactic disk. In this respect, \cite{Minchev18}
proposed a plausible way, for the first time, to obtain stellar birth radii
based on stellar age and metallicity with the help of assumed 
ISM evolution profile.
Recently, \cite{Feltzing19} applied this method to the APOGEE survey
and investigated the relative significance between the churning and blurring
processes by using different sets of ISM profiles.
It is of high interest to apply this method to open clusters
in the Galactic disk.
Since distances, metallicities, and particularly
ages of open clusters could be more reliably estimated as 
compared with stars in the field, more accurate birth sites
can be obtained and thus they become best tracers on the role of
radial migration in the Galactic disk.

The metallicity gradient has been well established for open clusters
but there is almost no age-metallicity relation \citep{Netopil16},
which is an evidence of the role of radial migration 
on open clusters.
In consistent with this, theoretical simulation by \cite{Martinez18}
provided strong arguments that NGC 6791
might have formed close to the inner disk of $3-5$ kpc or
the bulge, and strayed to the present location by
radial migration. The abundance analysis by \cite{Villanova18}
also supports its origin from the bulge population.
Meanwhile, it is found that %\cite{Quillen18} suggested that 
NGC 6583 could have radially migrated 3.5 kpc within
1 Gyr, indicating a high migration rate of 3.5 $\kG$ \citep{Quillen18}.
In this work, we aim to perform a systematic study
on a large sample of open clusters for the first time and probe the
role of radial migration on the Galactic disk
by estimating their migration distances
and measuring the migration rates, which could provide observational
constraints on theoretical mechanisms of radial migration.
It is expected that reliable ages of open clusters
will cast new sight not only on the evolution of the Galactic disk
but also on timescale of radial migration.

\section{Data and Sample Selection}
Although there are over 3000 open clusters in the Galaxy, only 
about 10\% have metallicity measurements in the WEBDA database by \cite{Dias02}.
These metallicities came from different sources and were derived
with different methods, leading to significant uncertainties and offsets. 
\cite{Netopil16} carefully checked these sources and provided a 
homogenous sample of 172 open clusters with metallicity (and also age)
adjusted to a consistent scale. This is our primary sample.
Meanwhile, \cite{Soubiran18} provided
accurate radial velocities for 861 open clusters
and calculated Galactic velocities based on these
radial velocities, the most probably distances
and proper motions in \cite{CantatGaudin18} from \textit{Gaia} DR2
\citep{Babusiaux18a,Brown18b}.
The primary sample from \cite{Netopil16} is 
cross-matching with the kinematic sample of
\cite{Soubiran18} and we obtain a sample of 148 open clusters.

Fig.~1 shows the metallicity gradient and the age-metallicity relation 
for this sample. Metallicities derived by different ways are marked
by different symbols: photometrical indices (126 open clusters; dots),  
high-resolution (80 open clusters; red open circles) and low-resolution
(39 open clusters; blue open circles)
spectroscopic data. It shows that metallicities from
high-resolution spectroscopic data have smaller scatters 
in metallicity
than those from low-resolution spectroscopic data and photometrical indices
at a given Galactic radial distance.
Among 148 open clusters in this
sample, we adopt metallicities from high-resolution spectroscopic data
for 80 open clusters, from low-resolution spectroscopic data for
10 open clusters and from photometric indices 
for the remaining 58 open clusters. 
The error in metallicity is about $0.06$ dex according to
\cite{Netopil16}.

The sample covers a metallicity range of $-0.50<\feh<+0.43$, 
an age range of $0.01<t<7.79$ Gyr and a Galactic radial distance 
range of $5<R<15$ kpc. 
%There is no clear trend in the age-metallicity relation.
Two exceptional open clusters, Berkeley 22 at 18.1 kpc and 
Berkeley 29 at 27.1 kpc, are excluded from this sample
because at such farther radii they do not follow the general
metallicity gradient (i.e. the decreasing $\feh$ trend 
with increasing $R$), which is required for estimating the birth site. 
Thus, our final sample includes 146 open clusters for further analysis.
This sample has advantage to probe the role of radial migration
at the early stage of the evolution of the Galactic disk because
the majority of open clusters are quite young.
Meanwhile, this sample has a wide span until 8 Gyr with 50 clusters 
older than 1 Gyr, and
thus it is still a good sample for probing the age effect of 
radial migration on the Galactic disk.

\begin{figure}
\includegraphics{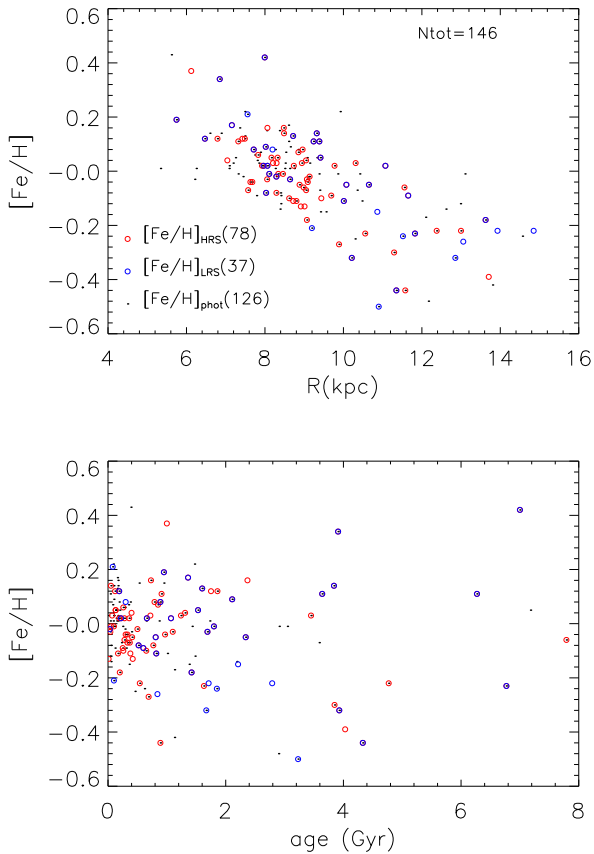}
\caption{The metallicity gradient and the age-metallicity relation for
of the selected sample of 146 open clusters. Metallicities from 
high-resolution spectroscopic data (78/80; red open circles), 
low-resolution spectroscopic data (10/39; blue open circles)
and photometric indices (58/126; black dots) are indicated seperately.}
    \label{fig:AMR}
\end{figure}

\section{Analysis}

\subsection{The birth site:$R_b$}
The birth sites are calculated based on the relation 
of $R_b-R_{\sun} = (\feh_{OC}- \feh_{ism}(R_{\sun},t))/gradient(t)$, 
where $t$ represents age of open cluster and $R_{\sun}$ is the solar radius
of 8.34 kpc \citep{Reid14} used in the present work.
The ISM metallicity at solar radius and the metallicity 
gradient as functions of age are taken from \cite{Minchev18}.
Ages and metallicities of open clusters by \cite{Netopil16} are 
adopted.
The error in the calculated birth site is based on uncertainties
of metallicity and age by the bootstrap method (1000 times for each cluster).
As shown in Fig.~2, the error distribution of the birth site 
has a peak at $\sim 0.1$ kpc with an extended tail toward $1$ kpc.
This is an internal uncertainty and it is expected that the external 
error could be larger. In Fig.~3, open clusters with $t<2.5$ Gyr 
at solar radius are compared with
the ISM profile of \cite{Minchev18}, and the metallicity gradient of very
young clusters with $0.1<t<0.5$ Gyr is of $-0.07\,dex/kpc$. 
It shows that there is no systematic
shift in the metallicity scale between our sample and the HARPS data of \cite{Minchev18},
although there is a substantial scatter in metallicity for young open clusters.
Meanwhile, the present gradient from very young clusters with $0.1<t<0.5$ Gyr is exactly
the same as the value used in \cite{Minchev18}.

\begin{figure}
\includegraphics{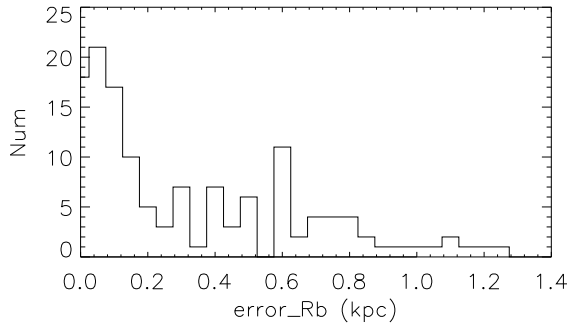}
    \caption{The error distribution of the birth site $R_b$ in this sample.}
    \label{fig:eRb}
\end{figure}

\begin{figure}
\includegraphics{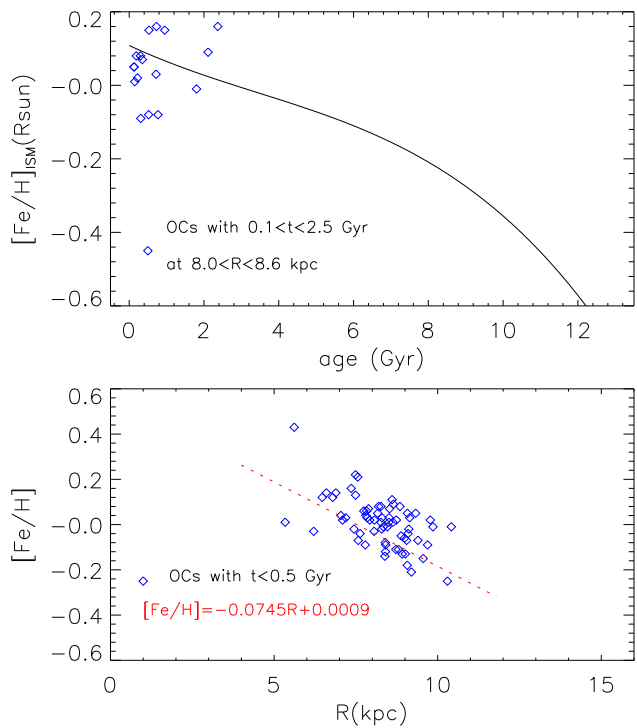}
    \caption{Upper: The evolution of ISM metallicity at solar radius
from Minchev et al. (2018) (black line) with
ovploted open clusters with $t<2.5$ Gyr at solar radius (blue diamonds).
Lower: The present metallicity gradient is of $-0.07\,dex/kpc$ (red dash line) dervied
from very young clusters with $0.1<t<0.5$ Gyr (black crosses).}
\label{fig:ISM}
\end{figure}

Fig.~4 shows the birth sites $R_b$ versus $\feh$
for five age ranges, $t<0.5$ Gyr, $t=0.5-1.0$ Gyr, $t=1.0-2.5$, $t=2.5-5.0$ Gyr and $t>5.0$ Gyr.
The birth site $R_b$ covers a wide range of $0-15$ kpc, as large as the whole $R$ range. Here, Galactic radial distance is of $R=\sqrt(X*X+Y*Y)$.
Four super metal rich ($\feh>0.2$) clusters come from the inner 
disk ($R_b<5$ kpc) and five metal poor ($\feh<-0.4$) clusters 
from the outer disk ($R_b>13$ kpc).
It seems that the main factor that determines $R_b$ is of metallicity.
 and the second factor of age.

In Fig.~5, the excursion distance, $(R-R_b)$, varies from $-4$ kpc to $+7$ kpc, and it
increases generally with age (and also with metallicity).
Specifically, $(R-R_b)$ is negative at $\sim-2$ kpc for young clusters
with $t<0.5$ Gyr, while it becomes positive with $(R-R_b)\sim+4$ kpc
for the oldest age bin of $t>5$ Gyr in our sample.
Five exceptional clusters, NGC 2243, Berkeley 99, Trumpler 5, Melotte 66 and
Berkeley 32, do not follow the general trend in the $(R-R_b)$ versus
age diagram. They have intermediate ages of $3-4$
Gyr but with the lowest metallicity of $\feh<-0.4$.

\begin{figure}
        \includegraphics{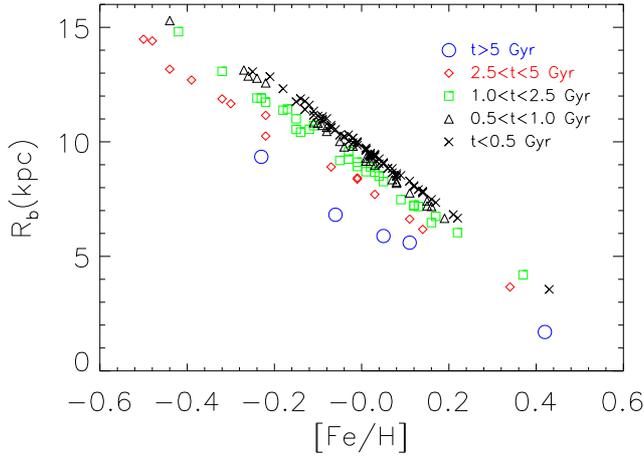}
    \caption{The birth site $R_b$ versus $\feh$
for five age intervals, $t<0.5$ Gyr (black crosses), $0.5-1.0$ Gyr (black triangles), $1.0-2.5$ Gyr (green squares),
$2.5-5$ Gyr (red diamonds) and $t>5$ Gyr (blue open circles).}
    \label{fig:RbFe}
\end{figure}

\begin{figure}
        \includegraphics{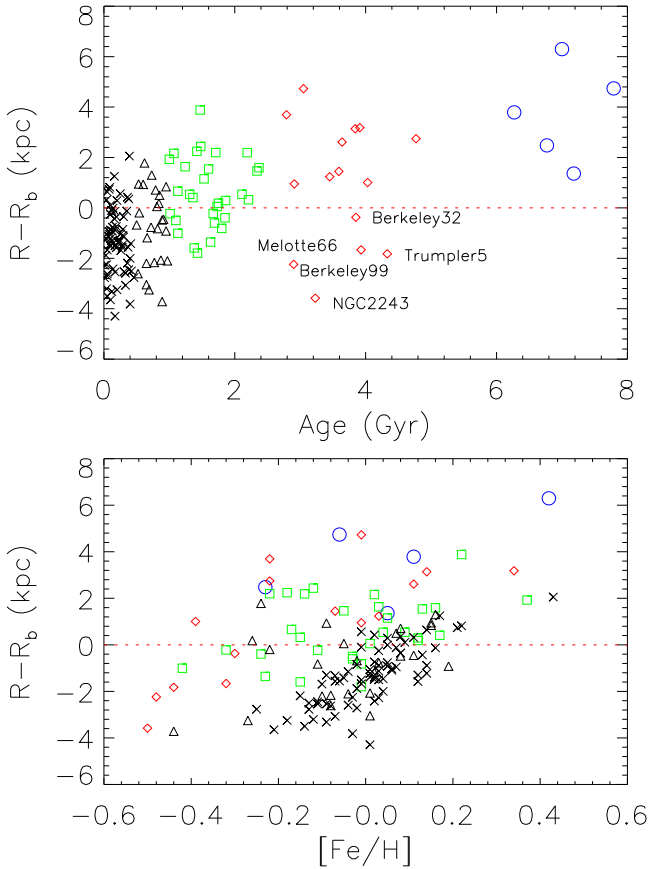}
    \caption{The excursion distance $R-R_b$ as functions of metallicity and age.The symbols are same as in Fig.~4.}
    \label{fig:RbMA}
\end{figure}

There is a weakly increasing trend of the excursion distance $(R-R_b)$ toward
the outer disk as shown in Fig.~6. 
This positive $(R-R_b)$ at $R>12$ kpc is in agreement with the 
results of \cite{Reddy16} who
suggested that the outer disc clusters were actually born inward
and radial migration had taken them to present locations.
Meanwhile, there is an age difference between the solar circle and
the outer disk clusters. Young clusters with $t<0.5$ Gyr covers a $R$ range 
of $6-10$ kpc but old open clusters 
span a wider R range from the inner disk toward the outer disk at $10-15$ kpc.
The relative number of old versus young clusters
increases with Galactic radius,
which indicates an increase of outward migrators toward the outer disk
where old clusters should be formed first and inside
(under the assumption of the inside-out formation) and then migrated to
the outer disk.
This supports the theoretical simulation work by
\cite{Minchev14}, who suggested that there should be an increase in
the relative number of outward migrators with Galactic radius.

\begin{figure}
        \includegraphics{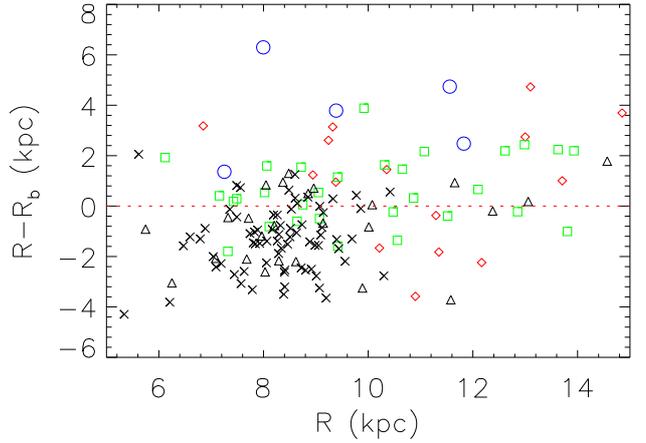}
    \caption{The excursion distance $R-R_b$ as functions of Galactic radial distance.
The symbols are same as in Fig.~4.}
    \label{fig:RbR}
\end{figure}

\subsection{Orbital parameters: $R_m$ and $R_g$}
The metallicity versus Galactic radial distance is shown
for the five age bins in Fig.~7.
It is clear that young open clusters with $t<0.5$ Gyr
have lower average $\feh$ than those with $0.5<t<2.5$ Gyr and $t>2.5$ Gyr.
According to \cite{Netopil16}, at $7<R<9$ kpc, the metallicity of old open clusters with
$1.0<t<2.5$ Gyr is 0.06 dex higher than that of young ones with $t<0.5$ Gyr
in their cleaned sample. They suggested that the increase of metallicity 
with age
can be explained by radial migration and this was supported by theoretical 
simulations of \cite{Minchev13} and \cite{Grand15}. 
That is, higher metallicity of old open clusters with $1.0<t<2.5$ Gyr
were born in the more metal rich inner regions and migrate outward
to present location. %, while young open clusters with $t<0.5$ Gyr
%were born locally, very close to their present locations.

\begin{figure}
        \includegraphics{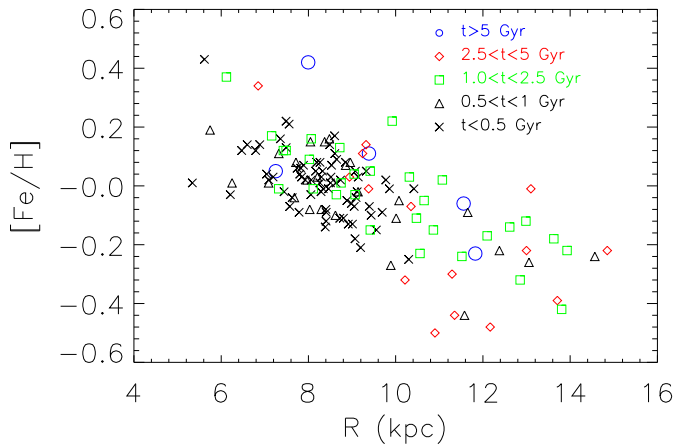}
    \caption{The $\feh$-$R$ for different age bins.The symbols are same as in Fig.~4.}
    \label{fig:RbMA}
\end{figure}

This direct comparison of average metallicity at a given Galactic location
does not work for older open clusters, i.e. $2.5<t<5.0$ Gyr and
$t>5.0$ Gyr, since there are only a few such clusters and 
they locate at different $R$ ranges. Instead, the comparison 
between the birth site and
orbital parameter provides a good way to evaluate the effect 
of radial migration.
For this purpose, we calculate the orbital parameters, peri-center/apo-center 
distances $R_p$, $R_a$ and guiding radius $R_g$
for the sample based on the axis-symmetric potential MWPotential2014
with the publicly available package \textit{Galpy} by \cite{Bovy15}. 
Here we
adopt the solar radius of 8.34 kpc and the circular velocity of 238 \kmprs
presented by \cite{Reid14}.

In order to investigate the blurring effect of the Galactic disk,
we show how different these orbital distances deviate from Galactic 
radial distance $R$ and how these deviations vary with Galactic locations.
Fig.~8 shows the $R_m-R$ and $R_g-R$ versus $R$,
where the mid-point distance is defined to be $R_m=(R_a+R_p)/2.0$.
For young open clusters with $t<0.5$ Gyr, $R_m-R \sim +1$ kpc and most
open clusters with $t<2.5$ Gyr locate within a limited range
of 1.5 kpc around the mean value of 1.0 kpc. 
But a few old open clusters at $R>10$ kpc have
large $R_m-R$ values ($3-5$ kpc).
Among 146 open clusters, 14 of them have $|R_g-R|>1$ kpc and they are 
found to have excursed at a distance of $\sim2$ kpc. 
That is, about 10\% open clusters have 
guiding radii quite different
from present locations as a result of the blurring process.
90\% open clusters have $|R_g-R|<1$ kpc, for which
radial excursions from their guding radii to present locations
are not significant.
Moreover, we note that almost all clusters with $t<2.5$ Gyr 
lie within $|R_g-R|<1$ kpc, while
clusters showing $|R_g-R|>1$ kpc are all older than 2.5 Gyr.
It seems that the blurring effect requires a timescale of 2.5 Gyr
to stray to a distance of larger than 2 kpc from the guiding radius.
Finally, we notice that these clusters locate mainly at $R> 10$ kpc,
which indicates that the blurring effect favors to work
outside the solar circle, e.g. at $10-15$ kpc.
In consistent with this suggestion, Fig.~9 shows that
open clusters undergoing large excursion
ranges of $R_a-R_p>5$ kpc also occur in the outer disk beyond 10 kpc.

\begin{figure}
        \includegraphics{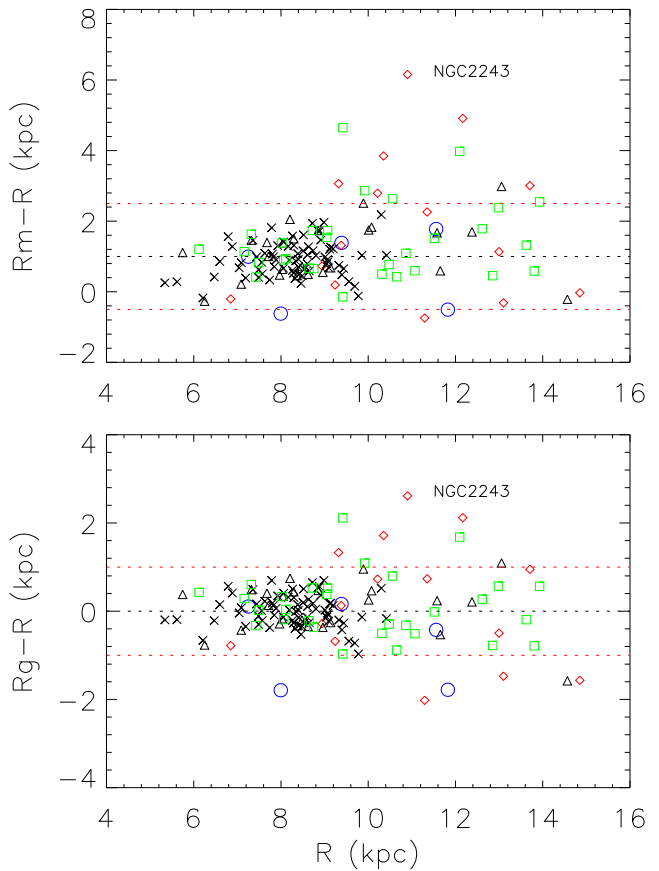}
    \caption{The $R_m-R$ and $R_g-R$ versus $R$ for different age bins.The symbols are same as in Fig.~4.}
    \label{fig:RbMA}
\end{figure}

\begin{figure}
        \includegraphics{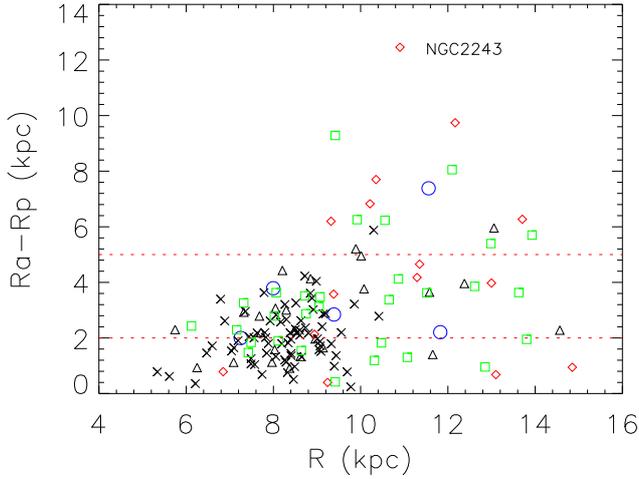}
    \caption{The $R_a-R_p$ versus $R$ for different age bins. The symbols are same as in Fig.~4.}
    \label{fig:RbMA}
\end{figure}

From Fig.~8, it seems that $R_g$ may be a better representative
for orbital distance than $R_m$ because $R_g$ is quite close 
to $R$ and $R_p$ for young open clusters
with $t<2.5$ Gyr, while $R_m$ shows a systematic shift 
of 1 kpc from $R$ even for very young open clusters with $t<0.5$ Gyr.
Moreover, since $R_g$ is directly related with the angular 
momentum by definition, it is a good proxy for current
orbital distance to investigate the churning effect of radial migration
in the following analysis.
When $R$ is replaced by $R_g$ in the X-axis of Fig.~7, it is
found, in Fig.~9, that the metallicity seems to decrease continuously with
$R$ beyond $R_g>12$ kpc, which makes it plausible
for us to estimate the
birth site for open clusters with $R>12$ based on a single
metallicity gradient.
The decreasing metallicity with $R$ was also found for Cepheid variables
by \cite{Luck11} and it is valid until $R\sim 15$ kpc, which
is the high limit of our sample. In addition, the dispersion of $\feh$
at a given $R_g$ is somewhat reduced for the outer disk beyond 12 kpc.
All these results show that $R_g$ is a good proxy for current
orbital distance.

\begin{figure}
        \includegraphics{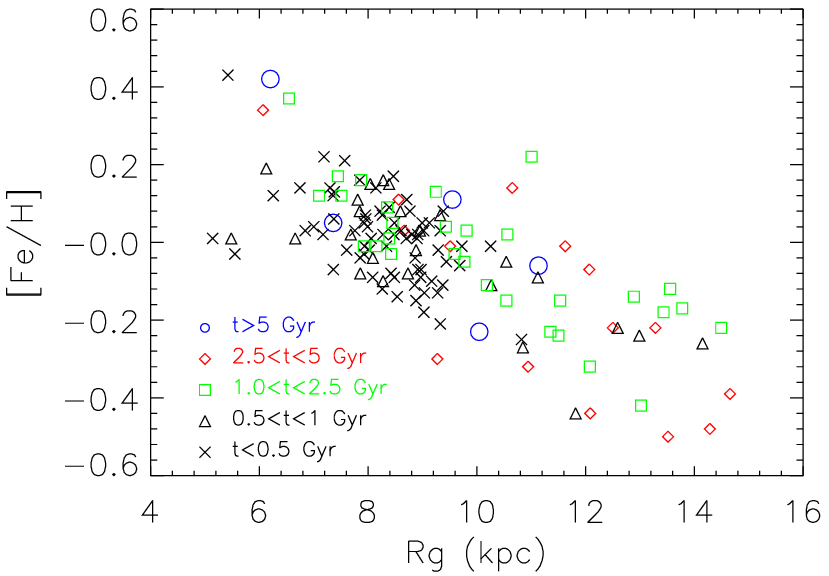}
    \caption{The $\feh$ versus $R_g$ for different age bins. The symbols are same as in Fig.~4.}
    \label{fig:RbMA}
\end{figure}

\subsection{The migration distance versus age diagram}
The migration distance, $R_g-R_b$, is a representative value
to estimate the effect of the churning process of
radial migration. Note that $R_b$ is caluculated
from metallicity and age, while $R_g$ is related with the kinematic
parameter, the circular velocity. Thus, the migration distance, $R_g-R_b$,
has the advantage to combine the metallicity, kinematics and age
information, which is important for tracing the evolution of the Galaxy.
Fig.~11 shows $R_g-R_b$ versus age for four $R$ ranges, $5<R<7$ kpc, 
$7<R<9$ kpc, $9<R<12$ kpc and $12<R<15$ kpc.
The upper panel is
for the whole age range and the lower panel shows the enlarged part
in the low age range of $t<2.5$ Gyr.
In general, young open clusters with $t<1.0$ Gyr
migrate inward with negative $R_g-R_b$ values, while older clusters 
with $t>1.0$ Gyr usually migrate outward with positive values. 
The increasing trend of $R_g-R_b$ with age is seen for young clusters
with $t<2.5$ Gyr and extends for the whole age range we investigated.
This provides a supportive evidence on the role of radial migration, 
which requires time for clusters to stray from their birth sites to 
the present positions, and thus older clusters tend to have larger
migration distances.

However, at all ages, there are open clusters with $R_g-R_b$
near zero (within 1 kpc); they do not experience significant churning process.
Thus, it needs a more specific inspect on how different radial migration
(or churning) works on individual clusters.
In the upper panel of Fig.~11, the black solid line separates open clusters 
move inward and outward, while the red solid line shows the migration distance
that can be reached during its lifetime
assuming a migration rate of 1 $\kG$ (see Chen et al. 2019 for details).
Taking into account of the maximum error of 1 kpc in $R_b$, we shift
the red solid upward by 1 kpc (the dash red line) and the two black dash
lines are drawn within 1 kpc around the black solid line. In this way, 
we define four
regions in Fig.~11: Region A above the red solid line, Region B below
the red line and above the upper black dash line,
Region C within the two black dash lines, and Region D below
the lower black dash line.

There are 16 open clusters at Region A with six of them
(NGC 6583, NGC 6603, Pismis 2, NGC 1798, Berkeley 73
and IC 166) above the red dash line. They move outward from their birth
sites to present locations, but the migration distance is larger 
than the reachable distance within their lifetimes under the assumption of
a migration rate of 1 kpc per Gyr. However, the migration rate may be
as large as 1.5 $\kG$ at the outer disk as we will show later, which
makes it easier for radial migration to account for clusters at Region A. 
Meanwhile,
we notice that the blurring effect may also play a significant role
for some clusters at this region. For example, IC 166 has $R_g-R=1.67$ kpc
and Pismis 2 has $R_g-R=1.09$ kpc.
Since most clusters at Region A locate within the red dash line,
it seems that radial migration generally accounts for
their origins.

Old open clusters with $t>2.5$ Gyr mainly locate at Region B,
where the well known old metal rich cluster, NGC 6791, is a typical case.
Due to their large ages, radial migration definitely accounts for 
the migration distances of these open clusters under the assumption 
of $1 \kG$ migration rate. The migration distance varies greatly
for old clusters, such as Collinder 261 and NGC 6791 as shown in Fig.~11. 
Thus, age is not
the only factor that determines the migration distance. 

Open clusters at Region C are defined to have migration distances around zero
(within 1 kpc); they do not experience significant radial migration.
Most clusters at Region C have ages less than $5.0$ Gyr with
only one exception (Berkeley 39). Berkeley 39 at $R=11.8$ kpc has $t\sim7$ Gyr
and the migration distance of $R_g-R_b \sim 1$ kpc, 
significantly lower than that of NGC 6791 at similar age.
The existence of Berkeley 39 
indicates that the outer disk has locally-born old cluster, which
is unusual under the assumption of the inside-out
formation scenario.

In Region D, most clusters are younger than 1 Gyr with 
one exception (Berkeley 32, 4 Gyr at $R=11.3$ kpc).
The migration distance in this region
covers a wide range of 4 kpc from $R_g-R_b=-5$ kpc to $-1$ kpc.
They locate within 12 kpc (with one exception, Berkeley 23 at $R=13.8$ kpc) 
where the metallicity gradient is valid. 
Finally, we note that the clusters from inner disk ($R<7$ kpc) have 
the largest span in migration distance
from the lowest to the highest values, although they are all
younger than 1 Gyr (with one exception, NGC 6253 with $t=3.91$ Gyr and $\feh=0.34$ dex). 

\begin{figure}
        \includegraphics{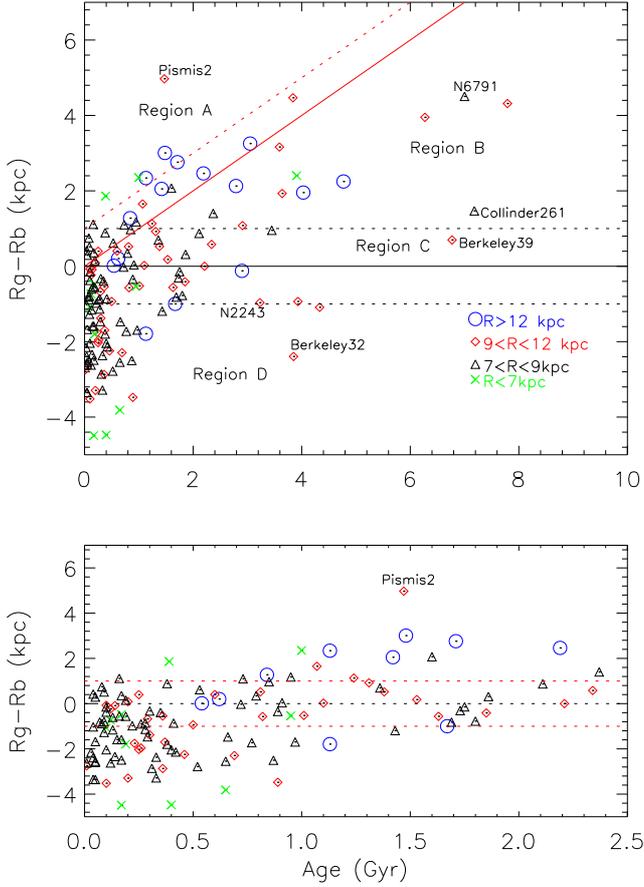}
    \caption{The $R_g-R_b$ versus age diagram for open clusters
at differnt $R$ ranges:
$5<R<7$ kpc (green crosses), $7<R<9$ kpc (black triangles),
$9<R<12$ kpc (red diamonds) and $R>12$ kpc (blue circles).
Lower panel shows the enalrged part of $0-2.5$ Gyr.}
    \label{fig:RgRb}
\end{figure}

\subsection{The migration rate}
With the migration distance and age, it is possible to estimate
a rough migration rate as defined to be the division of
the migration distance over the cluster's age.
Fig.~12 shows the distributions of the migration rate for open
clusters with $t>0.1$ Gyr as a function of age.
Very young open clusters with $t<0.1$ Gyr should be
excluded since very small value
in age would lead to a very large and uncertain migration rate.
For young clusters with $t<1.0$ Gyr, the migration rate
varies from $-5.0 \kG$ to $+2.5 \kG$, a wide range of 7.5 $\kG$.
The upper envelope
of migration rate is of $+2 \kG$ for intermediate-age clusters
with $1.0<t<3.0$ Gyr and it seems to decrease to $0.5-1.0 \kG$
for the oldest ones with $5<t<8$ Gyr.
In general, the histogram of the migration rate shows a peak at zero,
since 64 clusters (out of 146) in our sample do not experience 
radial migration with $|R_g-R_b| < 1 $ kpc. For clusters
moving outward ($R_g-R_b>1 $ kpc), the peak of the migration rate
is of $0.5 - 1.5 \kG$, while there is a wide distribution from $-5 \kG$
to $-1 \kG$ for inward-moving clusters with $R_g-R_b<-1$ kpc.

However, the above-defined migration rate might represent
a low limit for old clusters because we do not
know the exact time span of the migration process.
According to \cite{Minchev11}, radial migration
starts from $0.4$ Gyr and continues until $3.0$ Gyr 
or longer (see their Fig. 1), which
might indicate a time span of $\sim 2.5 - 3.0$ Gyr.
In view of this, the most plausible way to estimate the migration rate
in the present work
is based on intermediate-age outer-disk clusters, since
they have experienced a larger migration distance 
than those from the solar circle
and their ages are comparable to the time span of
the migration process.
Thus, we select open clusters with intermediate-age of $0.5 <t< 3.5$ Gyr and
at the outer disk of $12<R<15$ kpc and we obtain a migration rate of $\sim 1.5\pm 0.5 \kG$, as shown by the red solid line in the low panel of Fig. 12.
At the same age range, 
four clusters at the solar circle show a peak of migration rate 
at $\sim 1.0 \kG$.
These values are consistent or slightly larger than our adopted value 
of $1 \kG$ from \cite{Quillen18}, who
estimated the migration rate based on the Gaussian bar model by \cite{Comparetta12}.
This consistence shows the important role of the Galactic bar
in the theoretical modeling of radial migration process,
as already included in the chemical-dynamical model by \cite{Minchev13}.

\begin{figure}
        \includegraphics{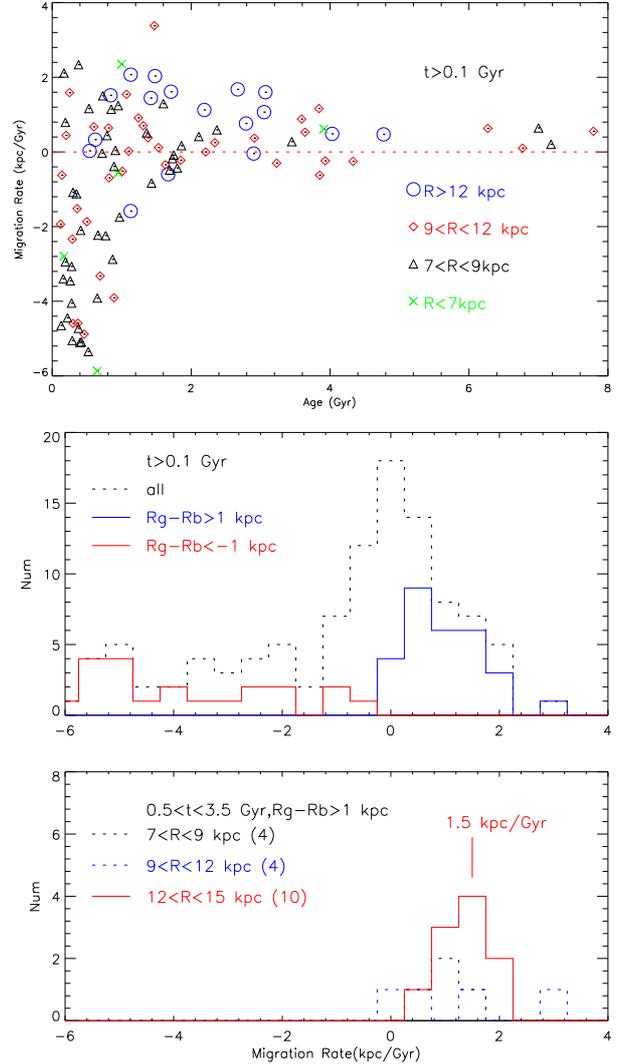}
    \caption{Upper: The estimated migration rate versus age
for all open clusters with $t>0.1$ Gyr (dash black line) at four $R$
ranges with the same symbols as in Fig.~11.
Middle: The histograms of the migration rate for migrated
clusters moving outward ($R_g-R_b > 1 $ kpc; blue solid line) and
inward ($R_g-R_b < -1 $ kpc; red solid line), as well as
for all clusters with $t>0.1$ Gyr (dash black line).
Lower: The migration rate is measured to be $\sim 1.5\pm0.5 \kG$
based on intermediate-age clusters ($0.5<t<3.5$ Gyr) at the outer disk
($12<R<15$ kpc). For comparison, the histograms from clusters 
at $9<R<12$ kpc (blue dash line) and $7<R<9$ kpc (black dash line)
are also shown.}
    \label{fig:Mrate}
\end{figure}

\section{Implications on the Mechanisms of Radial Migration and its
Role on the Galactic Disk}
Since open clusters belong to the young population of the Galactic disk,
it is expected that they do not suffer from significant radial excursions
caused by the blurring process, which is favored by the old population.
It is found that 90\% open clusters have $|R-R_g|<1$ kpc, which
indicates that the blurring effect do not
play a significant role for this young population of the Galactic disk.
However, only 44\% (64 out of 146) open clusters with $|R_g-R_b| <1$ kpc do not
suffer from significant churning process
as shown in the migration distance versus age diagram. 
The remaining 56\% clusters have
a wide range of absolute migration distance ($|R_g-R_b|$) 
from $1-5$ kpc for all ages. 
Those migrated clusters provide unique and valuable 
insights on the mechanisms of radial migration in the Galactic disk.

One of the most interesting result in the present work may be that the migrated
open clusters have different behaviors in migration direction.
53 clusters with $R_g-R_b<-1$ kpc
at Region D migrate inward, while 29 clusters with $R_g-R_b>1$ kpc
migrate outward, among which
13 can be explained by the churning process alone and
16 need other mechanisms working together.
As shown in Fig.~11, where the inward-migrated clusters with $R_g-R_b<-1$ kpc
are younger than 1.0 Gyr (the young group), while the outward-migrated 
clusters with $R_g-R_b>1$ kpc are older than 1.0 Gyr (the old group).
Moreover, the migration rates vary greatly in the young group and
there are more consistent values around $1-1.5 \kG$ for the old group.
Finally, the outward-migrated clusters mainly
belong to the outer disk with $R>9$ kpc (and dominate at $R>12$ kpc), while
the inward-migrated clusters have $R<12$ kpc and dominate
at the solar circle of $7<R<9$ kpc. 
These differences between the young and old groups among these migrated clusters
may indicate their different theoretical mechanicsms of radial migration.

The migrated clusters in the old group may be explained by the
coupled model of the bar and spiral arms of \cite{Minchev13}.
They show a relatively constant migration rate
of $\sim 1.5\pm0.5 \kG$, which is generally consistent
with the migration rate of 1 kpc/Gyr presented in \cite{Quillen18} for a pitch angle of 24
based on the Gaussian bar model by \cite{Comparetta12}.
Meanwhile, super metal rich stars
detected in the LAMOST survey by \cite{Chen19} also
favor the model by \cite{Minchev13}.
In this respect,
NGC 6791 is a typical case as the most metal-rich ($\feh=0.37$)
but quite old ($t=7.0$ Gyr) cluster \citep{Netopil16}.
It has thick disk kinematics locating at $R=6.1$ kpc and $|Z|=0.87$ kpc.
\cite{Linden17} suggested that it is unlikely
for a radial migration mechanism to operate by several kpc, and
especially to account for the current cluster's high altitude above
the plane. However, 
\cite{Martinez17} found that a fraction of the newly formed clusters
would eventually migrate more than 4 kpc and be lifted up to 0.8 kpc above
the disk's mid-plane. In their simulation, abundant examples of orbits
supported the scenario that NGC 6791 started
their evolution in the inner disk and reached to
the present location by radial migration.
Our results also suggest that radial migration accounts for 
its orbital parameters, migration distance and Galactic
location.

For the young group, Theoretical works by
\cite{Fujii12} might provide some hints on their migration mechanism.
According to their simulation, star clusters in non-steady disks
lose or gained angular momentum, and they could migrate inward
or outward by the order of a few kpc in a few hundred Myr.
The existence of very young clusters ($t<0.3$ Gyr) in Fig.12
provides observational support on this simulation.
Moreover, it seems that the migration rate is not
constant but varies greatly from $-6 \kG$ to zero.
But further works on the theoretical mechanisms, radial migration
or other scenarios, are necessary to explain the behaviors
of these very young open clusters with $t<1.0$ Gyr.

\section{Special Clusters at the Galactic Outer Disk}
The Galactic outer disk is significantly influenced by
both the blurring process
(10\% clusters with $|R-R_g|>1$ kpc showing $R_g>10$ kpc)
and the churning process (with a high migration rate of $1.5\pm0.5 \kG$).
Moreover, \cite{Quillen09} and \cite{Bird12} suggested that perturbations
caused by minor mergers are effective at mixing the outer disks
of galaxies.
Thus, it is expected that the Galactic outer disk has a complicate 
history, which might be seen from  some special clusters in the sample.

NGC2243 is one of the most special cluster in our sample
due to the lowest metallicity
of $\feh=-0.5$, the highest $Z_{max}$ of 2.65 kpc, the farthest apo-center
distance of 23.3 kpc and the highest eccentricity of 0.37. 
The high rotation of $\Vphi=278.3 \kmprs$
derived in this work and low $\alpha$ abundances 
from \cite{Francois13} indicate its
origin from the thin disk population. Its present location is far from
$R_g$, $R_m$ and has the largest excursion range of $R_a-R_p$ of 12.7 kpc.
The highest
eccentricity of 0.37 is consistent with the significant
effect of the blurring process,
while churning does not work as indicated by the $|R_g-R_b|<1$ kpc.
%The blurring effect seems to play an important role on this cluster.
The large $R_a=23.3$ kpc and high $Z_{max}=2.65$ kpc may
suggest that minor mergers also work in this intermediate-age cluster.
According to \cite{Minchev14}, the impact of mergers is the largest 
in the outer disks of galaxies due to the lower potential.

Pismis 2 has a typical thin-disk kinematics. The high metallicity 
of $\feh=0.22$ at the outer disk of $R_g=11.0$ makes it a special case,
since most stars at this radius is metal poor with $\feh\sim -0.4$.
The high metallicity indicates a small birth radius
from the inner disk, which leads to the largest
migration distance of $R_g-R_b=4.97$ kpc in our sample.
Based on its age of 1.47 Gyr, the migration
rate is around 3.4 $\kG$, the highest one among the old
group of migrated clusters. It is not consistent with
the values based on either other clusters at the outer disk ($1.5 \kG$).
The blurring effect is not significant with $R-R_g=1.08$ kpc, and
its eccentricity of 0.25 and low $Z_{max}=0.5$ kpc do not
favor for the scenario of minor mergers.
Further study on mechanisms of the churning process is desirable
to explain its high migration distance.

Berkeley 39 has a thick-disk kinematics, high $Z_{max}=1.36$ kpc and old
age of 8.0 Gyr. All of these indicates that it belongs to the thick disk 
population. The suggestion is further supported by its enhanced [$\alpha$/Fe]
ratios based on $[Mg/Fe]$, $[Al/Fe]$ and $[Si/Fe]$ abundances
derived from four giants by \cite{Friel10}.
It has a low eccentricity of $e=0.1$ and travels within 2 kpc 
between $R_p=10.2$ kpc and $R_a=12.4$ kpc. It is unlikely that
the blurring process or minor mergers works. Finally, $R_g-R_b$
is less than 1 kpc, which further excludes the
contribution from the churning process.
Thus, it is a locally-born old thick disk cluster at the outer disk.

In addition, there are other special clusters at the outer disk.
For example, Berkeley 23 (at $R=13.8$ kpc) is the only cluster 
at the outer disk 
that migrates inward. It has has $\feh=-0.42$ and $t=1.1$ Gyr with
a migration rate of 1.7 $\kG$, a typical value for the outer disk.
Its metallicity is typical at its Galactic radius and its young age
makes it special for the outer disk.
On contrary, Berkeley 32 at $R=11.2$ kpc is 4-Gyr-old cluster 
that migrated inward. Both the blurring and churning effects works together,
and its low eccentricity does not favor for the scenario of minor mergers.
The existence of different kinds of special clusters indicates
a complicated formation history of the Galactic outer disk.

\section{Conclusions}
\label{sect:conclusion}
A homogenous sample of 146 open clusters with available metallicity, age,
velocities and orbital parameters is obtained by cross-matching
the scaled metallicity/age catalog of \cite{Netopil16}
and accurate kinematic catalog of \cite{Soubiran18} based on Gaia DR2.
This sample spans a metallicity range of $-0.5$ to $+0.4$ dex,
a range in age of a few Myr to 8 Gyr and a range of 5 to 15 kpc
in radial radius. It allows us to explore the role of radial migration
on the Galactic disk by interpreting Galactic locations,
orbital parameters of open clusters and their migration distances 
as a function of age. 

The comparison of present location with orbital parameters, $R_m$ and $R_g$,
shows that 90\% open clusters have $|R_g-R|<1$ kpc and $|R_m-R|=1\pm1.5$ kpc.
Meanwhile, 10\% clusters with $|R_g-R|>1$ kpc and $R_m-R>2.5$ kpc mainly 
locate at the outer disk of $R>10$ kpc. This indicates that blurring 
causes significant radial excursions for only 10\% open clusters
in our sample, mainly at the outer disk.

The birth sites of open clusters are calculated, and
the deviation between the guiding distance and
the birth site ($R_g-R_b$) corresponds to the churning process of
radial migration. In our sample, 56\% open clusters suffer from 
significant churning and they move in two opposite directions.
Young clusters with $t<1.0$ Gyr migrate
inward by $2-5$ kpc, while old clusters with $t>1.0$ Gyr migrate
outward by $2-4$ kpc. The large migration distances of young
open clusters supports theoretical simulation
by \cite{Fujii12}, which suggested that the migration of several kpc could
happen within a few Myr. For older clusters, 
the chemo-dynamical evolution model by \cite{Minchev13} is the most
favorite mechanism, which successes to explain their distributions
in the migration distance versus age diagram.
Based on intermediate-age clusters at the outer disk, we measure the 
migration rate of $1.5\pm0.5 \kG$, which is consistent with the 
prediction of $1\kG$ by \cite{Quillen18} for a pitch angle of 24 based on
the Gaussian bar model by \cite{Comparetta12}.

Finally, we find several special clusters at the outer
disk, a thin-disk cluster (NGC\,2243) with the merging history 
excursing out to 23 kpc, a metal-rich outer disk cluster 
(Pismis 2) with the largest migration distance of 4.97 kpc,
and a locally-born old thick-disk cluster (Berkeley 39).
The existence of different kinds of special clusters 
implies a complicate formation history of the
outer disk of our Galaxy.

\section*{Acknowledgements}
We would like to thank the referee for his valuable suggestions, 
which greatly improve the manuscript.

This study is supported by the National Natural Science
Foundation of China under grant Nos. 11988101, 11625313, 11890694
and the National Key R \& D program of China  of 2019YFA0405502.

This work has made use of data from the European Space Agency (ESA) mission
{\it Gaia} (\url{https://www.cosmos.esa.int/gaia}), processed by the {\it Gaia}
Data Processing and Analysis Consortium (DPAC,
\url{https://www.cosmos.esa.int/web/gaia/dpac/consortium}). Funding for the DPAC
has been provided by national institutions, in particular the institutions
participating in the {\it Gaia} Multilateral Agreement.


%%%%%%%%%%%%%%%%%%%%%%%%%%%%%%%%%%%%%%%%%%%%%%%%%%

% Don't change these lines
\bsp	% typesetting comment
\label{lastpage}
\end{document}